\documentclass[12pt]{book}
\usepackage[dvips]{graphicx,color}
\usepackage{makeidx,tsukuba}
\makeauthorindex

\makeindex

\begin{document}
\BookTitle{\itshape The 28th International Cosmic Ray Conference}
\CopyRight{\copyright 2003 by Universal Academy Press, Inc.}
%\tableofcontents
\pagenumbering{arabic}

\chapter{%   %%%%%%%%% <===== TITLE of the contribution
%%%%%%%%%%% The first letter of each word should be capital letter.
Tau Neutrinos at EeV Energies}

\author{%
%
% You can include as many co-authors as you wish, unless
% the title/author information fits within 1 page.
%
S. Iyer Dutta,$^1$ I. Mocioiu,$^{2,3}$ M. H. Reno,$^4$ and I. Sarcevic$^2$ \\
{\it (1) Department of Physics and Astronomy, State University of New
York at Stony Brook, Stony Brook, New York 11794, USA \\ 
(2) Department of Physics, University of Arizona, Tucson, 
Arizona 85721, USA \\
(3) KITP, UC Santa Barbara, Santa Barbara, California 93106, USA\\
(4) Department of Physics and Astronomy, University of Iowa,
Iowa City, Iowa 52242, USA \\}
}%% end of author

\section*{Abstract}
Astrophysical sources of ultrahigh energy neutrinos yield tau neutrino
fluxes due to neutrino oscillations. At neutrino energies
in the EeV range, radio Cherenkov detection methods show promise
for detecting these fluxes. We quantify the tau neutrino contributions
to the signal in, for example, a detector like the Radio Ice
Cherenkov Experiment (RICE) for a $Z$-burst flux prediction.
Tau neutrino regeneration in transit through the Earth, including
energy loss, is evaluated.

\section{Introduction}

The experimental evidence of $\nu_\mu\leftrightarrow\nu_\tau$ neutrino
oscillations [5] means that astrophysical sources of muon neutrinos
become sources of $\nu_\mu$ and $\nu_\tau$ in equal proportions after
oscillations over astronomical distances [1,2]. Although $\nu_\mu$ and
$\nu_\tau$ have identical interaction cross sections, signals from
$\nu_\tau\rightarrow \tau$ conversions have the potential to contribute
differently from $\nu_\mu$ signals. The $\tau$ lepton can decay
far from the detector, regenerating $\nu_\tau$ [7-9]. This also occurs with
$\mu$ decays, but electromagnetic energy loss coupled with the long
muon lifetime make the $\nu_\mu$ regeneration from muon decays irrelevant
for high energies. A second signal of $\nu_\tau\rightarrow\tau$ is
the tau decay itself [3,4]. 

In this paper, we investigate the effect of $\nu_\tau$ regeneration from
tau decays in the neutrino energy range between $10^6-10^{12}$ GeV.
Attenuation shadows most of the upward-going solid angle for
neutrinos.
Regeneration comes from interaction and decay, so one is necessarily
considering incident neutrinos which are 
nearly horizontal or slightly upward-going in a discussion
of tau neutrino regeneration. The Radio Ice Cherenkov
Experiment (RICE) [11] has put limits on incident isotropic electron neutrino
fluxes which produce downward-going electromagnetic showers. Here, we
look at the $\nu_\tau$ contribution to horizontal
or upward-going electromagnetic showers where the shower is produced
in 1-4 km of ice. The Antarctic Impulsive Transient Antenna
(ANITA) also uses the ice as a neutrino converter [6]. The
ANITA balloon missions will monitor the ice sheet
for refracted radio frequency signals and require upward-going neutrino
interactions.

\section{Neutrino Attenuation and Regeneration}

Tau neutrino attenuation and regeneration is governed by its interaction
length and the tau decay length which is shown in Fig.\,1. We also plot 
the effective decay length with tau electromagnetic energy loss [10]. 
The neutral current neutrino(antineutrino) cross section
contribution to the total is about 1/2 of the charged current cross section.
As a comparison of the interaction lengths with physical distances 
we note that the chord (D) of the earth (in water equivalent distance) 
as a function of the nadir angle is given as
$D=2 R_\oplus\,\, \rho_{\rm rock}\,\,\cos\theta = 
5.9\times 10^7\ {\rm cm.w.e.}$ for
$\theta=89^\circ$.
Here  $R_\oplus=6.4\times 10^8$ cm, the mean Earth radius and
the density of standard rock $\rho_{\rm rock}=2.65$ g/cm$^3$.
Neutrino attenuation is clearly an important effect for nearly
horizontal neutrinos.

The effective decay length of produced taus does not go above $10^7$ cm,
even for $E_\tau=10^{12}$ GeV.
This is because electromagnetic energy loss over that distance reduces the
tau energy to about $10^8$ GeV, at which point, the tau is more likely 
to decay than interact electromagnetically [10].

\begin{figure}[t]
  \begin{center}
  \includegraphics[height=13.5pc,angle=270]{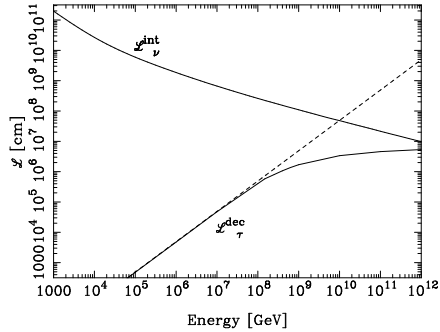}
  \end{center}
  \vspace{-0.5pc}
  \caption{Neutrino interaction length (solid line) and the tau effective
decay length without energy loss (dashed line) and with energy
loss in water (solid line).}
\end{figure} 

Attenuation and regeneration is governed by the neutrino transport equations,
as in, e.g., Ref. [8]. The coupled differential equations are solved
approximately using Euler's method for the neutrinos, which is modified
to include continuous tau decay for the taus.
Energy loss of the $\tau$ is accounted for by shifting
$E_{\tau,f}=E_{\tau,i}\exp(-\beta \Delta X)$ for column depth step
size $\Delta X$ and $\beta$ from the
average high energy loss formula $-dE_\tau/dX\simeq \beta E$.
As a first approximation, we use $\beta=0.8\times 10^{-6}$ cm$^2$/g.

\section{Results}

\begin{figure}[t]
  \begin{center}
  \includegraphics[height=11.0pc,width=16.5pc]{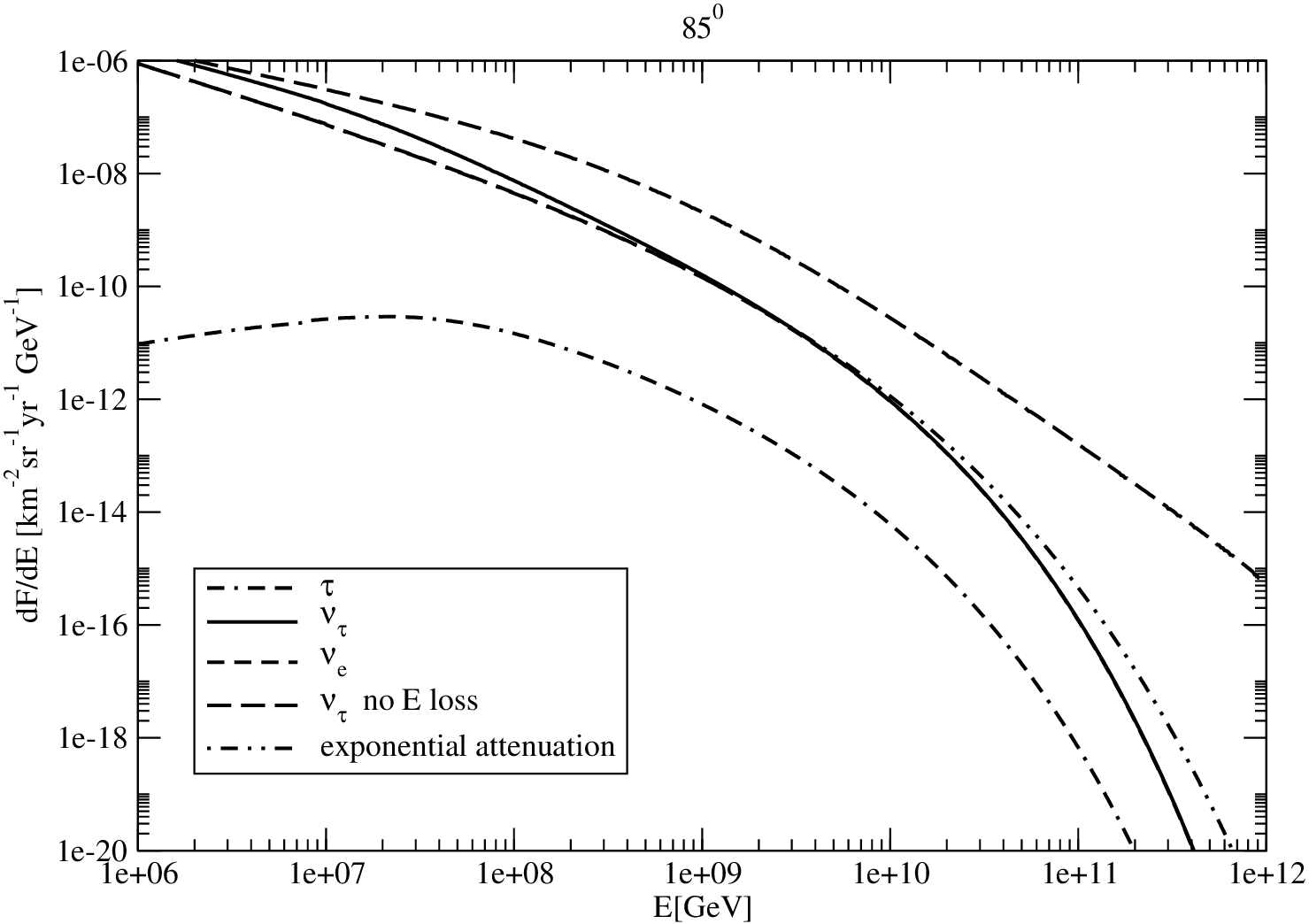}
  \end{center}
  \vspace{-1.3pc}
  \caption{Electron neutrino, tau neutrino and tau fluxes for an initial
$Z$-burst flux at a nadir angle of 85$^\circ$.}
\end{figure}

Fig.\,2 shows the attenuated/regenerated $\nu_e$ and $\nu_\tau$ fluxes
as well as the $\tau$ flux for the $Z$-burst [12] model of 
Ref. [13], approximated by
\begin{equation}
{dF_\nu\over dE}\, 
[{\rm km^{-2}yr^{-1}sr^{-1}GeV^{-1}}]= \left\{ \begin{array}{ll}
1/(E/{\rm GeV})\quad& \mbox{$E<2.5\times 10^{12}$ GeV}\\
6.24\times 10^{24}/(E/{\rm GeV})^3\quad & \mbox{$E\geq 2.5\times 10^{12}$
 GeV}
\end{array}
\right.
\end{equation}
and we use a nadir angle $\theta=85^\circ$.
We approximate the Earth density over the course of the trajectory to be
$\rho_{\rm rock}$ [$D=2.9\times 10^8$ cm.w.e.]. Little
$\nu_\tau\rightarrow \tau\rightarrow \nu_\tau$ regeneration occurs, as can
be seen by comparing the $\nu_\tau$ flux (solid line) with the $\nu_e$
flux (dashed line just below solid curve). 
The dot-dot-dashed line shows the result of simple attenuation with
$\exp (-D/{\cal L}^{\nu}_{CC}(E))$, which agrees with the
$\nu_e$ flux for $E
{\mathrel{\raisebox{-.6ex}{$\stackrel{\textstyle<}{\sim}$}}} 10^{10}$ GeV. 
The uppermost dashed line 
corresponds to the tau neutrino flux without the energy loss.
The tau flux (dot-dash-dashed line) at the end of the
neutrino trajectory through the Earth is a factor of $\sim 10^{-5}-10^{-3}$
of the attenuated tau neutrino flux for the energy $E=10^{6}-10^{11}$ GeV.

In Fig.\,3, we show the electron fluxes from $\nu_e$ charged current
(CC) interactions and from $\tau\rightarrow e$ for the $Z$-burst model,
again at $\theta=85^\circ$, for two different column depths in water:
$D=1$ km and 4 km. In evaluating the fluxes, we have integrated
over energy distributions of the tau decay or the neutrino interaction
cross section. The most important contributions of $\tau\rightarrow e$ to
the electromagnetic signal occur below $10^8$ GeV. For taus, electromagnetic
energy loss and the falling probability to decay in depth $D$ suppress
$\tau\rightarrow e$ high energy contributions. For $\nu_e$, the rising
cross section works in the opposite direction.

We conclude that the electron contribution to the electromagnetic signal
from $\tau$ decay from nearly horizontal incident $\nu_\tau$ is a small
portion of the $\nu_e\rightarrow e$ CC signal, except in the energy
range of $\sim 10^6-10^8$ GeV, for the
$Z$-burst flux model. For the RICE experiment, this is in the
energy range where their effective detection volume is smallest, making
the $\tau$ signal from $Z$-burst tau neutrinos difficult to extract.

\begin{figure}[t]
  \begin{center}
  \includegraphics[height=11.0pc,width=16.5pc]{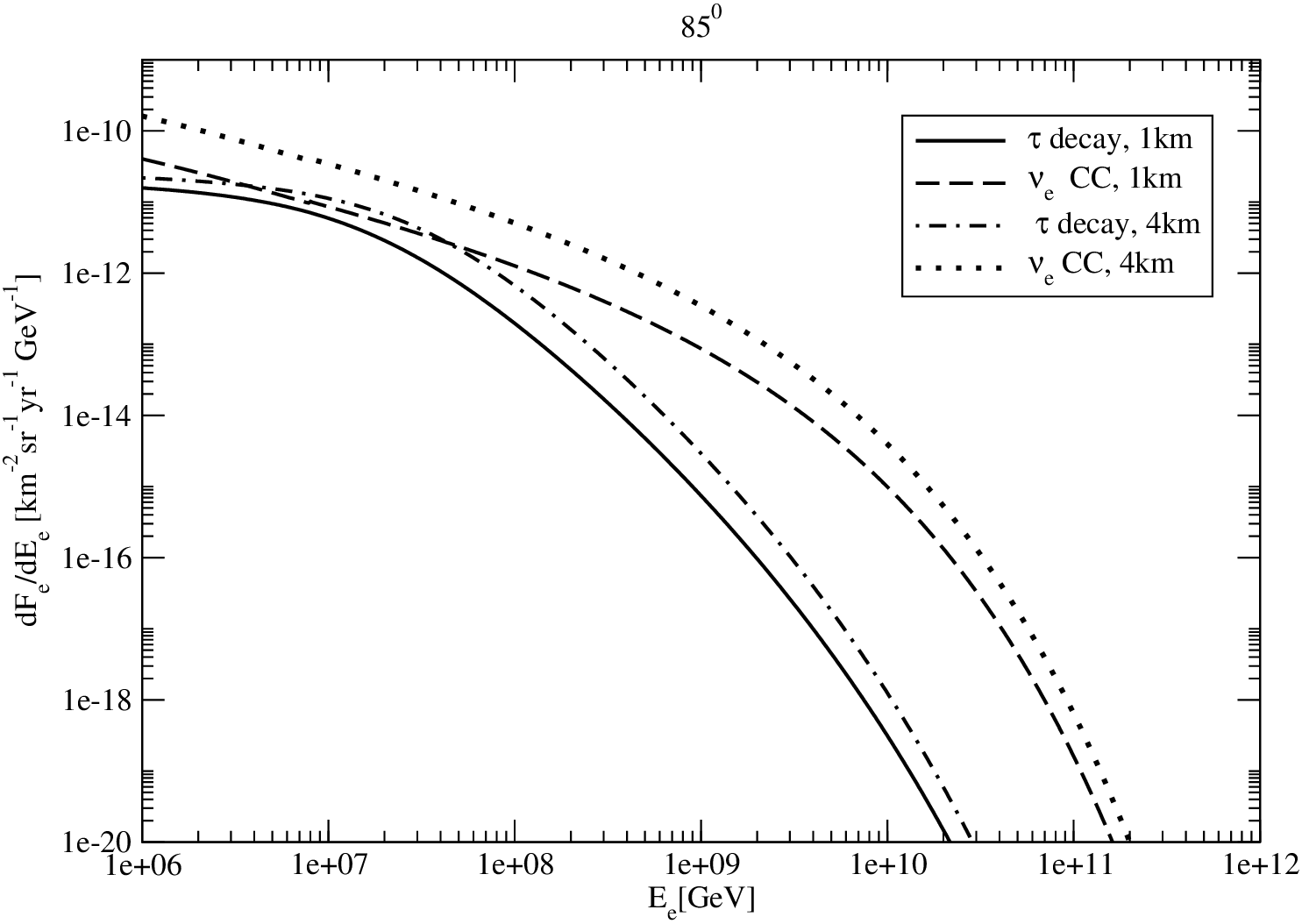}
  \end{center}
  \vspace{-1.5pc}
  \caption{Electron fluxes from $\tau\rightarrow e$ and 
$\nu\rightarrow e$ given an initial
$Z$-burst flux for $\theta =85^\circ$.}
\end{figure}

\section{Acknowledgments}

The work of S.I.D. has been supported in part by NSF Grant No. 0070998.
The work of I.S. and I.M. has been supported in part by the DOE under
contracts No. DE-FG02-95ER40906 and DE-FG02-93ER40792, and NSF Grant
No. PHY99-07949. The work
of M.H.R. has been supported in part by the DOE under contract number
DE-FG02-91ER40664. I.S. and I.M. thank the KITP at UC Santa Barbara
for hospitality. We thank J. Jones and J. Sayre for discussions.

\section{References}

\re 
\ 1.\ Ahluwalia D.V., Ortiz C.A., Adunas G.Z. 2000, hep-ph/0006092 
(unpublished)
\re
\ 2.\ Athar H., Jezabek M., Yasuda O. 2000, Phys. Rev. D 62, 1033007
\re
\ 3.\ Fargion D. 2002, Ap. J. 570, 909
\re
\ 4.\ Feng J.L., Fisher P., Wilczek F., Yu T.M. 2002, Phys. Rev. Lett. 88,
161102
\re
\ 5.\ Fukuda Y. et al. [Super-Kamiokande Collaboration] 1998, Phys. Rev.
Lett., 81, 1562
\re
\ 6.\ Gorham P. et al. [ANITA Collaboration]\hfil\break
http://www.ps.uci.edu/$\sim$%
barwick/anitaprop.pdf
\re
\ 7.\ Halzen F., Saltzberg D. 1998, Phys. Rev. Lett. 81, 4305
\re
\ 8.\ Iyer S., Reno M.H., Sarcevic I. 2000, Phys. Rev. D 61, 053003
\re
\ 9.\ Iyer Dutta S., Reno M.H., Sarcevic I. 2000, Phys. Rev. D 62, 123001
\re
10.\ Iyer Dutta S., Reno M.H., Sarcevic I., Seckel D. 2001, Phys. Rev. D 63,
094020
\re
11.\ Kravchenko I. et al. [RICE Collaboration], astro-ph/0206371
(unpublished)
\re
12.\ Weiler T.J. 1999, Astropart. Phys. 11, 303
\re
13.\ Yoshida S., Sigl G., Lee S. 1998, Phys. Rev. Lett. 81, 5505

\endofpaper
\end{document}